\documentstyle[aps,preprint]{revtex}

\begin{document}

\title{Locating Transition States Using Double-ended Classical Trajectories}

\author{A. Matro and D. L. Freeman}

\address{Department of Chemistry, University of Rhode Island,
Kingston, Rhode Island 02881}

\author{J. D. Doll}

\address{Department of Chemistry, Brown University,
Providence, Rhode Island 02912}


\maketitle

\begin{abstract}
In this paper we present a method for locating transition states and
\mbox{higher-order} saddles on potential energy surfaces using
\mbox{double-ended} classical trajectories.  We then apply
this method to 7- and 8-atom
\mbox{Lennard-Jones} clusters,
finding one previously unreported
transition state for the 7-atom cluster and two for the 8-atom
cluster.
\end{abstract}
\pacs{PACS numbers: }

\pagebreak

\section{Introduction}
Classical dynamics calculations in multi-dimensional systems
are most commonly carried out by numerically propagating Newton's
equations of motion.  This approach requires that initial momenta of the
particles be specified as well as initial coordinates.  Although this
approach has been invaluable in numerous applications such as Molecular
Dynamics simulations, there are certain problems in which
the use of double-ended boundary conditions
can be a more convenient choice.

In the double-ended formulation of the dynamics, the initial and
the final coordinates of the system are specified rather than the
initial coordinates and momenta.  Trajectories that connect the initial
and the final configurations of the system are then determined.  Doll,
Beck and Freeman \cite{dbf} proposed an approach to obtain double-ended
trajectories by using the Fourier expansion about the linear
constant-velocity path connecting the two configurations.
Double-ended classical trajectories
were recently constructed by Cho, Doll and
Freeman \cite{cho} to investigate the dynamics of the Henon-Heiles potential.
Both bound and dissociative trajectories were located for this
potential.  In other recent work, Gillian and Wilson have used a
discretized formulation of double-ended trajectories to study the
dynamics of isomerization and reactive collisions.\cite{wilson}
In related work, Pratt has outlined a method based on Monte Carlo
sampling for locating transition states using the reactant and product
configurations as the boundary conditions in the construction of reactive
trajectories. \cite{pratt}

Our goal in this paper is to use the double-ended trajectory
approach
to find transition states connecting a pair of potential energy
minima.
The use of double-ended trajectories
seems natural in the search for transition-state geometries since
an initial and a final configuration of the system are specified
explicitly.
Although there might exist many trajectories that
connect a pair of potential energy minima in a given transit time,
 it is possible to identify
those trajectories that take the system through or close to an
extremum on the potential energy surface.  In this paper we will refer to
trajectories that pass through a transition state
as {\em transition-state trajectories}.
Furthermore, it is possible to
restrict the search procedure to favor the low-energy trajectories that
are likely to be transition-state trajectories.

Other methods previously employed in transition state searches include
the method developed by Cerjan and Miller.\cite{cerjan}
This method uses a procedure
that maximizes the potential energy of the system with respect to
displacement of one normal mode while simultaneously minimizing the
potential energy with respect to displacements of the remaining normal
modes.  The system, therefore, follows a particular normal mode until a
transition state is reached.  In recent work, Davis {\em et al.} \cite{davis}
and Wales \cite{wales} have employed
this method as well as the method of slowest slides \cite{berry}
in locating transition states in a
wide class of systems including atomic Lennard-Jones and ionic
clusters.

This paper is organized as follows. Section 2 reviews the formalism
behind the construction of double-ended classical trajectories and
describes the criteria for identifying transition-state trajectories.
Because the transition-state trajectory for a given transit time
 must be the lowest energy
trajectory connecting a pair of potential energy minima,
the computational procedure is tailored to locate simple
low-energy trajectories.  In Section 3, we use this method to locate
transition states between isomers of atomic Lennard-Jones clusters.
For the 7- and 8-atom clusters we are able
to locate previously unreported low-lying transition state(s).
Concluding remarks are given in Section 4.

\section{Method}

The expressions in this section are presented for a one-dimensional
system to keep the notation simple.
Extension to three dimensions and many particles is straightforward.
The path traversed by a particle
between points $x_1$ and $x_2$ in time $t$ (referred to from now on
as {\em transit time}) can be written using a
Fourier sine expansion about the constant-velocity path,\cite{dbf,cho}
\begin{equation}
x(u) = x_1 + (x_2-x_1)u+\sum_{k=1}^\infty a_k \sin (k\pi u) \; ,
\label{eq:xofu}
\end{equation}
where $u\equiv \tau/t$ is the dimensionless time ranging from 0 to 1,
$\{a_k\}$ is the set of
expansion coefficients that describes the deviation of the trajectory
from a constant-velocity path, and $\tau$ is the physical
time along the trajectory.
The expansion coefficients $\{a_k\}$ may be
obtained by substituting Eq. (\ref{eq:xofu}) into Newton's equations of
motion or equivalently by requiring that the action be stationary with
respect to variations
of the
expansion coefficients. The resulting coefficients are given by
\cite{dbf,cho}
\begin{equation}
a_k = -\frac{t^2}{m\pi^2k^2} f_k({\bf a}) \; , \label{eq:asubk}
\end{equation}
where $m$ is the mass of the particle
and $f_k({\bf a})$ is the $k$th Fourier sine
component of the force on the particle, given by
\begin{equation}
f_k({\bf a}) = 2\int_0^1 du \sin (k\pi u) \mbox{\Large(} \!-
\frac{dV(x(u))}{dx(u)}\mbox{\Large)}  \; .
\label{eq:fsubk}
\end{equation}
The boldface ${\bf a}$ in Eqs. (\ref{eq:asubk}) and (\ref{eq:fsubk})
implies that each component $f_k({\bf a})$ generally depends
on the entire set of expansion coefficients.
The coefficients $a_k$ can be determined by finding the
zeros of the function
\begin{equation}
\chi({\bf a}) = \sum_{k=1} \delta_k^2 \mbox{\Large [}a_k+
\frac{t^2}{m\pi^2k^2}f_k({\bf a})\mbox{\Large ]}^2
\; , \label{eq:chiofa}
\end{equation}
where $\delta_k$ are arbitrary real constants.  In a previous
application\cite{cho} it was convenient to introduce the weighting
factors, $\delta_k$, to move between regions of $\chi({\bf a})$ that
correspond to
different paths.  In the current work, we have set $\delta_k=1$.
Each zero of the function $\chi({\bf a})$ corresponds to a classical path
that can be traversed by the system between $x_1$ and $x_2$ in the specified
transit time.

The strategy used by Cho, Doll and Freeman in locating classical
trajectories on a Henon-Heiles potential consisted of randomly choosing
values for the finite set of expansion coefficients and then minimizing
$\chi({\bf a})$ given in Eq. (\ref{eq:chiofa}). \cite{cho}
Their goal was to find
all possible trajectories for a given transit time.  As the transit time was
increased, larger numbers of coefficients
were required in the Fourier expansion
of the trajectory to describe the increasingly complicated
motions undertaken by the particle.

It is not our goal in this work to determine all
possible trajectories that take the
system from one configuration to another. Instead, we wish to restrict our
search to yield only simple low-energy trajectories that can take
the system through or close to an extremum on the potential energy
surface.
The transition states of atomic Lennard-Jones clusters have been
recently studied in detail \cite{davis,wales} and we choose these systems
to test our method.
The Lennard-Jones potential is given by
\begin{equation}
V = 4\epsilon \sum_{i<j} \mbox{\Large [(}\frac{\sigma}{r_{ij}}
\mbox{\Large )}^{12} -
\mbox{\Large (}\frac{\sigma}{r_{ij}}\mbox{\Large )}^6\mbox{\Large]}
\; , \label{eq:ljpot}
\end{equation}
where $\epsilon$ is the parameter governing strength of the interaction,
$\sigma$ determines the pair equilibrium separation, and $r_{ij}$
is the distance between atoms $i$ and $j$.

The strategy that we employ here consists of
using a small number of expansion coefficients for each degree of freedom
to exclude the possibility of complicated trajectories
from our search. We are looking for the most direct
trajectory between the two points on the potential energy surface. The
family of trajectories on which we focus is
analogous to the trajectories giving rise to the
left-most branch in the energy-time diagram for
dissociation on the Henon-Heiles potential in Fig. 2 of reference \cite{cho}.
In addition, transitions from one local minimum to another often
involve motion of only a few atoms in the cluster,
because most rearrangements in small systems can be
described with either of two mechanisms: the Johnson edge-bridging
mechanism \cite{johnson} or the Lipscomb diamond-square-diamond mechanism.
\cite{lipscomb} Of course the kinds of rearrangements in large many-body
systems
can be complex and collective, requiring an analysis beyond the scope of
the current work.
The simplification described above allows us to further restrict our
search by randomly
choosing the
expansion coefficients for the coordinates of only those atoms that move
significantly during the rearrangement, while the rest of the expansion
coefficients are initially set to zero.  All the coefficients are, of
course, optimized in the search for the zeros of
$\chi({\bf a})$.  The minimization of $\chi({\bf a})$ is performed
using the conjugate gradient method. \cite{recipes}
For all the transition state
searches carried out in this work, three expansion coefficients for each
Cartesian coordinate are used and it is only necessary to randomly
choose the $k=1$ coefficients for the atoms whose positions change
significantly during the rearrangement.
The range used to generate the initial values of the
randomly chosen expansion coefficients is varied from $\pm 0.1\sigma$ to
$\pm 1.0\sigma$ depending on the particular pair of potential energy minima.

In order for the search to be most efficient, the relative
orientations for each
pair of minima must be carefully chosen so that as few atoms as possible
move during the rearrangement.
Figure 1 shows the relative orientation
of a pair of potential energy minima of the 7-atom cluster that yields the
transition-state trajectory efficiently. Note that the only atom that
moves significantly is atom 4. Proper numbering of the
atoms in the two clusters is important as well, since inconsistent
numbering of atoms in the two structures will force a complicated
rearrangement.

A parameter that must be chosen with some care is the transit
time.  The transit time cannot be too short, since that would
result in large kinetic energies, leading to trajectories that are not
very sensitive to the underlying potential energy surface (see the
short-time behavior of the energy-time diagram in Fig. 2 of reference
\cite{cho}).
If the transit time is chosen to be too short, the kinetic energy
required to complete the trajectory will greatly exceed the potential
energy of the cluster, resulting in ballistic trajectories.
On the other
hand, if a transit time is chosen to be too long, then there is more
opportunity to find an undesirable (i.e. complicated) trajectory.
We have found that using
a transit time that is about 8-10 times longer than the
vibrational period of a Lennard-Jones dimer to be most optimal in
determining the transition-state trajectories for Lennard-Jones clusters.

Finally, we need to distinguish between those trajectories that pass
through or close to the transition state and those that do not.
A transition-state trajectory will have the following properties:
a) the total energy of a transition-state trajectory must be the
lowest of all the trajectories found between a particular pair of local
minima; and  b) the kinetic energy of the system will just be barely
enough to allow the system to go over the barrier.  From these two properties
we see that the potential
energy plotted as a function of time will be nearly constant for a
significant portion of transit time as the system slowly moves through
the transition state. Figure 2 shows the plot of the potential energy as
a function of the dimensionless time $u$
for a transition-state trajectory (thin curve)
and a non transition-state trajectory (thick curve)
between the two minima shown in Figure 1.
Both curves begin and end at the potential energy corresponding
to the structures shown in Figs. 1a and 1b, respectively.  The
curve corresponding to the transition-state trajectory is
dominated by a prominent
plateau-like region throughout a large part of the trajectory.  The
large zero-slope portion of this curve suggests that the cluster
spends a significant portion of its trajectory in a region of the
potential energy surface where the forces vanish, i.e. near an extremum.
If the cluster is given barely enough kinetic energy to surmount the
barrier, it will move extremely slowly near the
extremum.
As the transit time is increased, the size of the plateau-like region
of the time-dependent
potential energy curve increases as well.
The thick curve in Fig. 2 exhibits three maxima and no prominent plateau-like
region as seen in the thin curve.  There is no indication in the appearance
of the
thick curve in Fig. 2 that the cluster passes through a
potential energy extremum at any stage of its trajectory.

It must be emphasized that the method described above does not determine
the transition state geometry precisely.  Two factors are responsible for
this.
First, since we are attempting to describe the dynamics
of the system with a finite number
of Fourier coefficients, we are not determining the true dynamics of
the system.
Second, the system is required to pass through a transition
state (or an extremum) only in the limit of an infinitely long transit
time.
To locate the transition state precisely, we pick a
configuration in the transition-state (plateau) region of the trajectory and
minimize the sum of
the squares of the gradients of the potential.  Usually, the
potential energy of the
initially chosen configuration
is within 0.05$\epsilon$ of the actual potential energy
of the transition state.
The transition state is confirmed by diagonalizing the Hessian matrix
and verifying that there is one negative eigenvalue corresponding to the
reaction coordinate, along with six zero
eigenvalues corresponding to rotations and translations and a
remaining set of positive eigenvalues corresponding to the bound normal modes.

\section{Results}

\subsection{Seven-Atom Cluster}

We have used the double-ended classical trajectory method to locate
transition states between potential energy minima of the seven-atom
Lennard-Jones cluster.
The seven-atom cluster has four minima, two of which have been
shown in Fig. 1.
Previous studies by Davis {\em et al.} \cite{davis}
found eight transition states connecting these four minima with the use of
the slowest-slide \cite{berry} and Cerjan-Miller \cite{cerjan} methods.
In testing our method, we have located all previously
reported transition states among the
isomers of the 7-atom Lennard-Jones cluster.  All trajectory searches
were
carried out with three expansion coefficients for each Cartesian
coordinate.  Since the presence of a plateau-like region in the plot of
the potential energy along the trajectory can
only suggest an extremum on the potential energy
surface and not necessarily a transition state, we did not know
{\em a priori}
whether a trajectory was in fact a transition-state trajectory.  Once a
possible transition-state trajectory was identified, the geometry of
the cluster in the plateau-like region of the trajectory
was optimized by minimizing
the sum of the squares of the gradients.  Once this sum was sufficiently
small, the eigenvalues of the Hessian matrix were evaluated to
determine whether the extremum was a transition state or
a higher order saddle point.

We have also located a transition state not reported previously.
The structure of this transition state is shown in Figure 3. This
transition state that connects the tricapped tetrahedron with itself
(see Fig. 1c of reference \cite{davis}), has a potential energy of
-14.57$\epsilon$
and is similar
to the transition state shown in Fig. 2h of reference \cite{davis}
previously found that connects the bicapped trigonal
bipyramid with itself (see Fig. 1d of reference \cite{davis}).
In both cases, the transition state occurs via the
edge-bridging mechanism. \cite{johnson}

In addition
to the transition states, we have also found a second-order saddle point
in the potential energy surface between the lowest-energy
isomer, the pentagonal bipyramid, and the tricapped tetrahedron. What
makes this particular saddle point, shown in Fig. 4,
interesting is the fact that its
potential energy of -14.75$\epsilon$ is lower than that of two
previously reported
transition states shown in Figures 2g and 2h of
reference \cite{davis} and the transition state shown in Fig. 3 of this paper.
This new saddle point can be expected to
contribute to the low-energy dynamics of the 7-atom cluster
on an equal footing with energetically similar transition states.

\subsection{Eight-Atom Cluster}

A number of transition states among the four lowest-energy
isomers of the eight-particle
Lennard-Jones cluster have been found by Wales.\cite{wales}
We found two previously unreported
transition states between the capped pentagonal bipyramid and the
bicapped octahedron. The structures for these minima are shown
in Figures 3a and
3c of reference \cite{wales}, respectively, and their potential
energies are -19.83$\epsilon$ and -19.19$\epsilon$.
The transition states, shown in Figs. 5a and 5b,
both have lower energies (-18.71$\epsilon$ and -18.68$\epsilon$, respectively)
than the previously reported transition state between
these two minima (-18.42$\epsilon$, shown in Fig. 4d of reference
\cite{wales}).

The search for the transition states of the 8-atom cluster also
revealed a limitation of our method.  We were unable to locate a
transition-state trajectory between the dodecahedron and
the bicapped octahedron, shown Figures 4b and 4c of reference \cite{wales},
respectively.
The lowest-energy trajectory that we found connecting these two
structures did not yield a potential energy along the trajectory
characteristic of a transition-state trajectory.
This curve, shown in Fig. 6, does not exhibit the characteristics of
the lower (thin) curve in Fig. 2.  Further
investigation of geometries along the trajectory corresponding to the
plot in Fig. 6 revealed that the configuration of the cluster at
a time corresponding to the small shoulder-like feature in Fig. 6 (at
$u \sim .25$) was
close to the previously reported transition state,
shown in Fig. 4b of reference \cite{wales}.
The transition-state trajectory should, therefore, have its maximum
potential energy at that configuration.

In an effort to understand the reason behind the inability of our method
to locate the transition state between the bicapped octahedron and
the dodecahedron we conducted a search for trajectories connecting each
of the two potential energy minima with the transition state.  Ideally,
the potential energy along the lowest-energy trajectory from a
potential energy
minimum to a transition state should increase monotonically.  This
was the case for the trajectory originating at the bicapped octahedron.
A trajectory with monotonically increasing potential energy
starting at the dodecahedron, however,
could not be located. The lowest-energy trajectory that was found
between the dodecahedron and the transition state passed through
configurations with potential energies above that of the transition
state.

To study this case in more detail we ran molecular dynamics
trajectories (using standard coordinate-momentum boundary conditions)
starting at configurations slightly displaced from the
transition-state geometry along the reaction coordinate.  The atoms
in the clusters
were given no initial momenta.   When the configuration was displaced
toward the bicapped octahedron, the cluster immediately found the
corresponding potential energy minimum.  On the other hand, when the
starting geometry was displaced toward the dodecahedron, the lowest
potential energy reached by the cluster before a subsequent increase was
{}~0.4$\epsilon$ above the corresponding potential energy minimum.  The
results of the standard molecular dynamics trajectories together with the
lack of
a transition-state trajectory between these two minima imply
that a dynamical transition-state trajectory
connecting these two minima does not exit.
More specifically, the transition
state is not dynamically accessible from the minimum corresponding to
the dodecahedron with a trajectory whose total energy does not exceed the
potential energy at the transition state.  This result implies that a careful
search for transition states along the lowest energy trajectory is
warranted even when the lowest energy trajectory does not have the
characteristics of a transition-state trajectory.

\section{Concluding Remarks}
We have presented a novel approach for locating transition states in a
multi-dimensional potential.  The double-ended
classical trajectory approach has been adapted to search efficiently for
trajectories that take the system from one potential energy minimum to
another through or close to an extremum on the potential energy
surface.
With this technique, we have been able to
locate a previously unreported transition state for the 7-atom
Lennard-Jones cluster and two previously unreported transition states
for the 8-atom cluster.

We have also tested our approach on the 13-atom cluster and were able
to locate several of the transition states reported in reference
\cite{wales}.  We did not conduct an extensive search for the transition
states of the 13-particle cluster because we were mainly
interested in determining the applicability of our method to larger
clusters.  The typical times for determining the set of expansion
coefficients for a trajectory ranged from under half a minute for the
7-atom cluster to about 2 minutes for a 13-atom cluster on a Silicon
Graphics Indigo workstation.  The computing time scaled as the square
of the number of atoms and, therefore, significantly larger systems are
certainly accessible with this method.

The appeal of our method is the ability to use physical intuition in
choosing the relative orientations and numbering of atoms in a pair of
potential energy minima between which a transition state is being sought.
This way one can force the cluster to undergo a specific rearrangement
and determine whether a transition state exists for such a path.
We have also learned that a successful application of this method
relies on the existence of a dynamical transition-state trajectory.  As
it turned out, a transition state for the 8-atom cluster was
not dynamically accessible from one of the minima without passing
through configurations with potential energies above that of the
transition state.  This implies the need for a careful search of the
lowest energy trajectory to seek any hidden transition states.

\section*{Acknowledgments}
Acknowledgement is made to the Donors of the Petroleum Research Fund of
the American Chemical Society for support of this work.  This work was
also supported by NSF Grant No. CHE-9203498.  Figures 1,3,4 and 5 were
drawn using XMol, version 1.3.1, Minnesota Supercomputer Center, Inc.,
Minneapolis, MN, 1993.

\pagebreak
\section*{Figure Captions}
\noindent
Figure 1.  Local minima of the 7-atom Lennard-Jones cluster:
a)pentagonal bipyramid $-16.51\epsilon$,
b)capped octahedron $-15.94\epsilon$.
\\
\\
Figure 2.  Potential energy of 7-atom cluster along the trajectory
connecting structures shown in Figs. 1a and 1b.  Lower (thin) curve
corresponds to a transition-state trajectory, and upper (thick) curve
is for a higher-energy trajectory.
\\
\\
Figure 3.  Transition state between two tricapped octahedra,
$V = -14.57\epsilon$.
\\
\\
Figure 4.  Geometry at the second-order saddle point between the
pentagonal bipyramid and the capped octahedron, $V = -14.75\epsilon$.
\\
\\
Figure 5.  Transition states for the 8-atom cluster between the capped
pentagonal bipyramid($-19.83\epsilon$)  and the bicapped octahedron
($-19.19\epsilon$): a) $-18.71\epsilon$, b) $-18.68\epsilon$.
\\
\\
Figure 6.  Potential energy along the lowest-energy trajectory for the
rearrangement between the dodecahedron and the bicapped octahedron
minima of the 8-atom cluster.


\begin{references}
\bibitem{dbf}
J. D. Doll, T. L. Beck and D. L. Freeman, {\em Int. J. Quant. Chem.}
Quantum Chemistry Symposium {\bf 23}, 73 (1989).
\bibitem{cho}
A. E. Cho, J. D. Doll and D. L. Freeman, Chem. Phys. Lett, in press.
\bibitem{wilson}
R. E. Gillian and K. R. Wilson, {\em J. Chem. Phys.} {\bf 97}, 5713
(1992).
\bibitem{pratt}
L. R. Pratt, {\em J. Chem. Phys.} {\bf 85}, 5045 (1986).
\bibitem{cerjan}
C. J. Cerjan and W. H. Miller, {\em J. Chem. Phys.} {\bf 75}, 2800
(1981).
\bibitem{davis}
H. L. Davis, D. J. Wales and R. S. Berry, {\em J. Chem. Phys.} {\bf 92},
4308 (1990).
\bibitem{wales}
D. J. Wales {\em J. Chem. Phys.} {\bf 91}, 7002 (1989).
\bibitem{berry}
R. S. Berry, H. L. Davis and T. L. Beck, {\em Chem. Phys. Lett.}
{\bf 147}, 13 (1988).
\bibitem{johnson}
B. F. G. Johnson, {\em J. Chem. Soc., Chem. Commun.} {\bf 1986}, 27
(1986).
\bibitem{lipscomb}
W. N. Lipscomb, {\em Science} {\bf 153}, 373 (1966).
\bibitem{recipes}
W. H. Press, S. A. Teukolski, W. T. Vetterling and B. P. Flannery, {\em
Numerical Recipes}, (Cambridge University Press, Cambridge,
1986), Ch. 10.
\end{references}
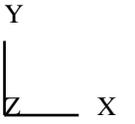
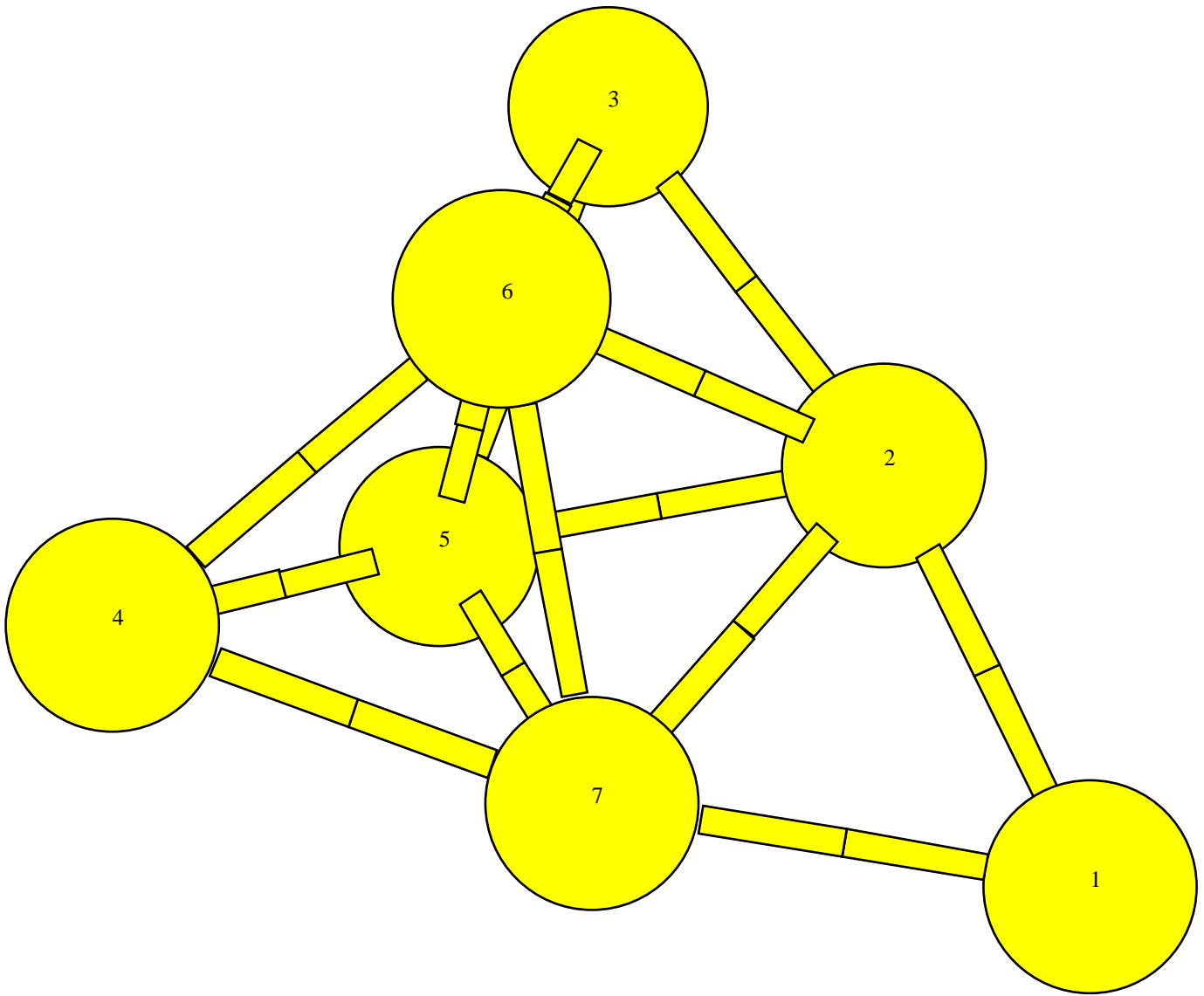
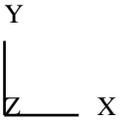
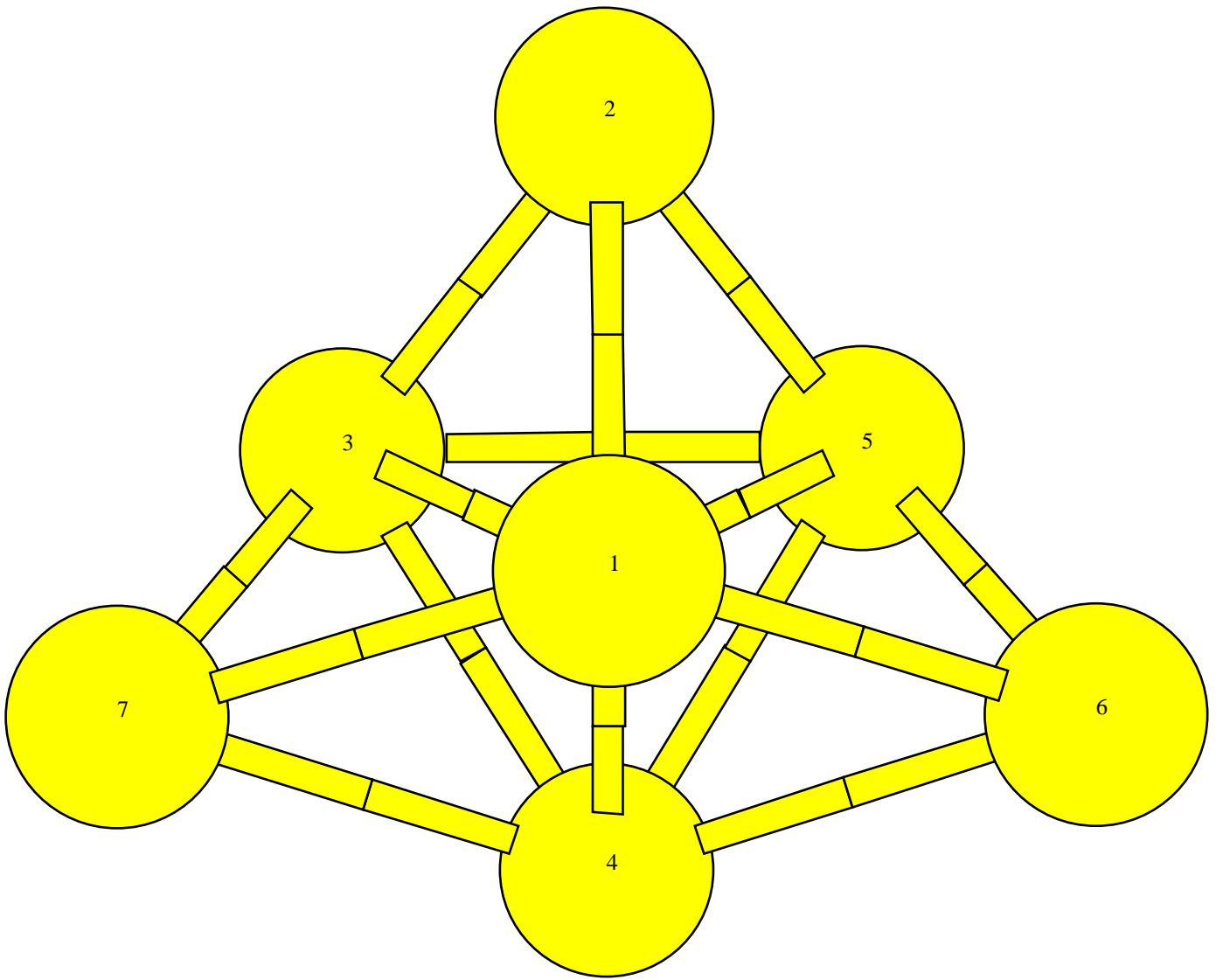
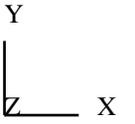
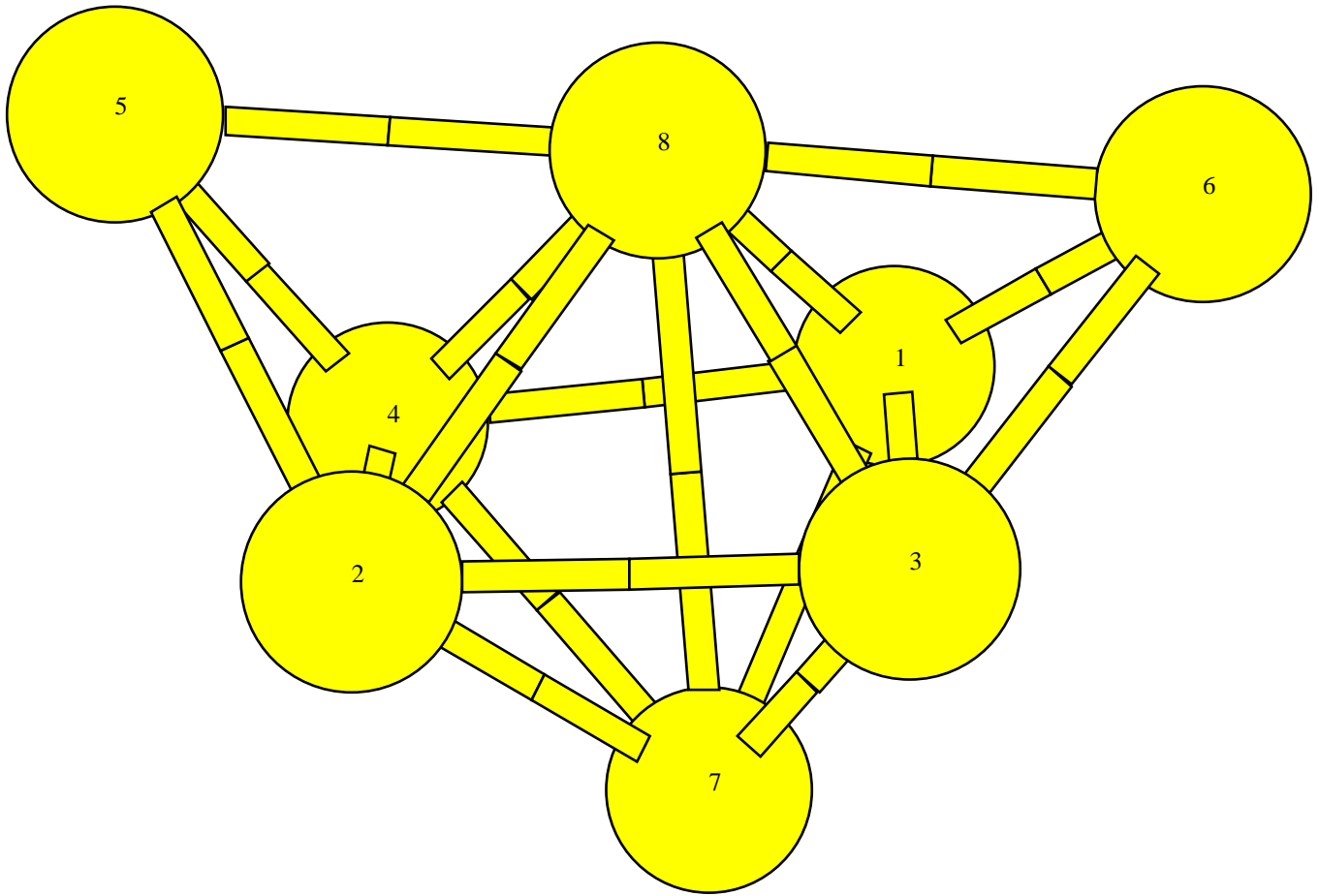
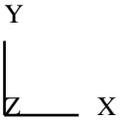
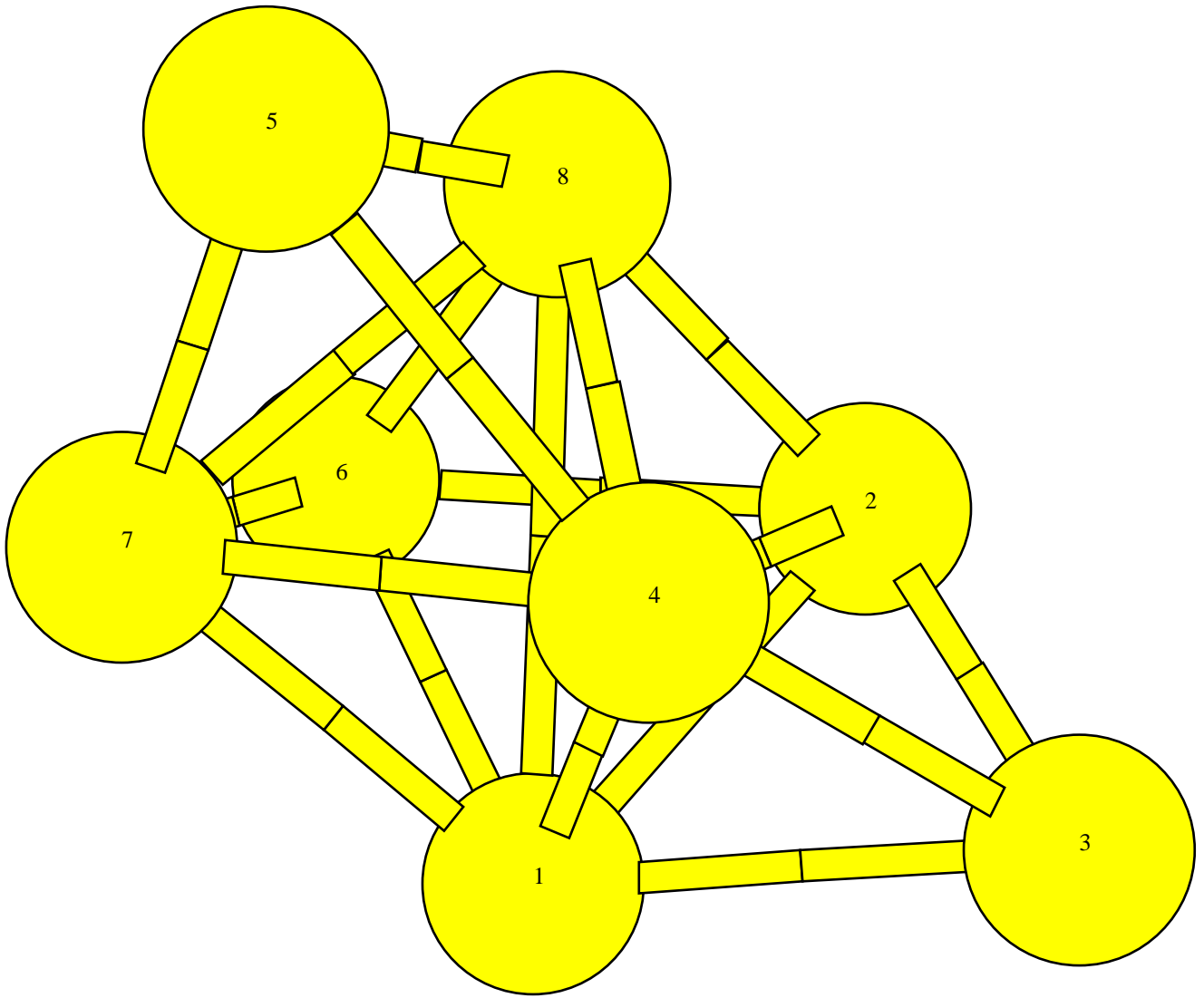
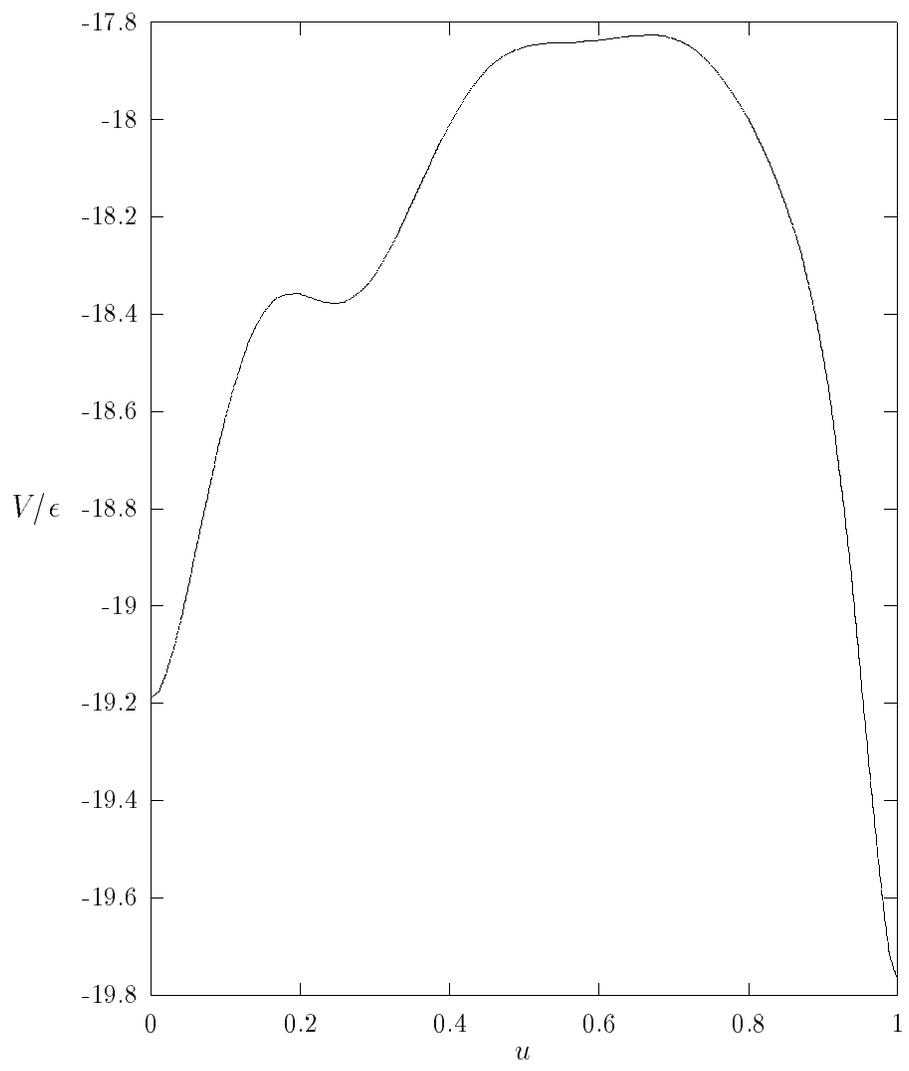
\end{document}